\documentclass[aps,prb,showpacs,showkeys,superscriptaddress,twocolumn,10pt]{revtex4-1}
\usepackage{amsmath,graphicx,color
}
\usepackage{dcolumn}
\usepackage{bm}

\def\suma{\sum\displaylimits}

\begin{document}
\title{The 30-band {\bf k}$\cdot${\bf p} modeling of electron and hole states in silicon quantum wells}
\author{Nemanja A. \v{C}ukari\'c}\email{nemanja.cukaric@etf.bg.ac.rs}
\affiliation{School of Electrical Engineering, University of
Belgrade, P.O. Box 35-54, 11120 Belgrade, Serbia}
\affiliation{Department of Physics, University of Antwerp,
Groenenborgerlaan 171, B-2020 Antwerp, Belgium}

\author{Milan \v{Z}. Tadi\'c}\email{milan.tadic@etf.bg.ac.rs}
\affiliation{School of Electrical Engineering, University of
Belgrade, P.O. Box 35-54, 11120 Belgrade, Serbia}

\author{Bart Partoens}\email{bart.partoens@ua.ac.be}
\affiliation{Department of Physics, University of Antwerp,
Groenenborgerlaan 171, B-2020 Antwerp, Belgium}

\author{F. M. Peeters}\email{francois.peeters@ua.ac.be}
\affiliation{Department of Physics, University of Antwerp,
Groenenborgerlaan 171, B-2020 Antwerp, Belgium}
\date{\today}
\begin{abstract}
We modeled the electron and hole states in ${\rm Si}/{\rm SiO}_2$  quantum wells within a basis of standing waves using the 30-band {\bf k}$\cdot${\bf p} theory. The hard-wall confinement potential is assumed, and the influence of the peculiar band structure of bulk silicon on the quantum-well sub-bands is explored. Numerous spurious solutions in the conduction-band and valence-band energy spectra are found and are identified to be of two types: (1) spurious states which have large contributions of the bulk solutions with large wave-vectors (the high-$k$ spurious solutions) and (2) states which originate mainly from the spurious valley outside the Brillouin zone (the extra-valley spurious solutions). An algorithm to remove all those nonphysical solutions from the electron and hole energy spectra is proposed. Furthermore, slow and oscillatory convergence of the hole energy levels with the number of basis functions is found and is explained by the peculiar band mixing and the confinement in the considered quantum well. We discovered that assuming the hard-wall potential
leads to numerical instability of the hole states computation. Nonetheless, allowing the envelope functions to exponentially decay in a barrier of finite height is found to improve the accuracy of the computed hole states.
\end{abstract}


\maketitle

\section{Introduction}

Silicon remains the technologically most important
semiconductor, and is used to build numerous
electronic\cite{lauhon2002,xiang2006} and photonic
devices.\cite{steigmeier1996,boucard2004,yang2006,michelini2011}
For almost four decades, the scaling down of the dimensions of these devices has been driven into the current nanoscale by Moore's law. Therefore, modeling transport and optical properties of silicon nanodevices should take into account quantum-confinement effects. In silicon quantum wells, the valence-band states can be modeled by the six-band {\bf k}$\cdot${\bf p} theory,\cite{kurdi2003} whereas, the conduction-band states require application of more elaborate  {\it ab initio}\cite{peelaers2006,petretto2011,blase2006,markussen2007,wu2008} or tight-binding\cite{proot1992,nishida1998,allan1997,delerue2000,lansbergen2011} methods because of the indirect band gap nature of silicon. However, the latter calculations become increasingly complex for larger structures, leading to slow performance and the requirement for large computer memory. These conditions are alleviated by the {\bf k}$\cdot${\bf p} theory, which can successfully describe states close to the band extrema. Yet, those {\bf k}$\cdot${\bf p} Hamiltonians usually are for states close to the $\Gamma$ point of the Brillouin zone,\cite{sheng2002,rastelli2004,seguin2005} and are, therefore, suitable for nanostructures made of direct band-gap materials.

The approach, which has recently been pursued as a successful
alternative to atomistic calculations, is the 30-band {\bf k}$\cdot${\bf p} model.\cite{cardona1966,rideau2006,fraj2008,saidi2010} This model was demonstrated to accurately describe states in the whole Brillouin zone and was applied to silicon nanostructures\cite{richard2005,kurdi2006} and ${\rm GaAs}/{\rm AlGaAs}$ superlattices.\cite{even2011}  Nevertheless, because
of the large number of energy bands which are taken into account, the 30-band calculations are far from being trivial. Also, this Hamiltonian is not invariant with respect to translational symmetry of the crystal, which is a general drawback of the {\bf k}$\cdot${\bf p} theory. Moreover, the results of {\bf k}$\cdot${\bf p} models can exhibit spurious solutions,\cite{foreman1997,kolokolov2003,yang2005,zhao2012} which  arise from the incorrectly determined  bulk states and, therefore, represent a considerable hurdle for calculations of the electron and hole states in a nanostructure.  As a matter of fact, some of the envelope functions of the spurious solutions are highly oscillatory. A way to avoid them is to cut off contributions of the bulk states with a large wave vector.\cite{yang2005} The use of a basis consisting of plane waves is a natural choice for such a method, as recent calculations have demonstrated for InGaAs/InP superlattices.\cite{yang2005} Also, spurious solutions might also arise due to the lack of ellipticity of the multiband Hamiltonian.\cite{veprek2007,eissfeller2011}. Unfortunately, no general method for removing spurious solutions from the {\bf k}$\cdot${\bf p} calculation has been proposed to date. Furthermore, almost all the proposed methods for the removal of spurious solutions were tested on the eight-band {\bf k}$\cdot${\bf p} Hamiltonian and for direct-band gap semiconductors.

In this paper, we study the electronic structure of silicon quantum wells by the 30-band {\bf k}$\cdot${\bf p} model.\cite{rideau2006} We consider ${\rm Si}/{\rm SiO}_2$ quantum wells grown along the [001] direction. Because the conduction- and valence-band offsets are quite large, the electrons and holes are mainly confined in the silicon layer. Therefore, for convenience, an infinite potential-well confinement (hard-wall potential) was assumed, and the basis of standing waves was used. The conduction-band states of this quantum well have recently been considered by the approximate effective two-band model.\cite{michelini2011} The values of the parameters were taken from Ref.~\onlinecite{rideau2006}, where the dispersion relations of the bulk bands in the whole first Brillouin zone (FBZ) were fitted to the results of {\it ab initio} calculations. However, the symmetry between the FBZ and the second Brillouin zone (SBZ) was not established in this fitting procedure. Therefore, spurious solutions are found in the energy spectrum. We explore their origin and, moreover, formulate a procedure which removes them from the energy spectrum of the analyzed quantum well. Also, the stability of some of the solutions with respect to variation in the order of the basis is discussed. Moreover, to explore, in more detail, how the specific boundary conditions affect the solutions, we supplement the analysis for the case of the finite band offset between Si and ${\rm SiO_2}$. In all our results, the top of the silicon bulk valence band is taken as the zero of energy.

\begin{figure}
    \begin{center}
       \includegraphics[width=8cm]{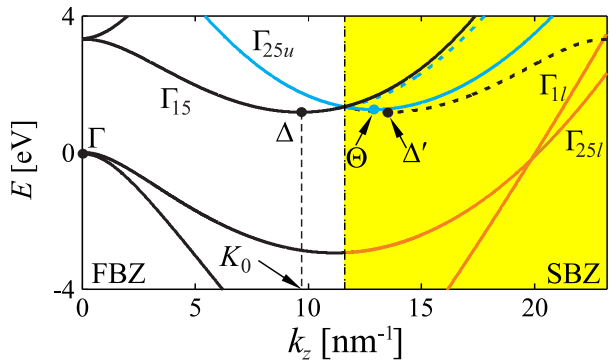}
       \caption{\label{fig1}(Color online) The bulk band structure of silicon along the $[001]$
       direction. The vertical dot-dashed line is the boundary of the first Brillouin zone. The extra valley which arises from the $\Gamma_{25u}$ band is denoted by $\Theta$, and the dashed curves denote the energy bands with the proper symmetry.}
    \end{center}
\end{figure}

\section{The bulk band structure of silicon}

Before presenting the results of our calculations for the modeled
silicon quantum well, we briefly discuss the silicon bulk band
structure, as computed by the 30-band {\bf k}$\cdot${\bf p} model.\cite{rideau2006} The dispersion relations of a few bands with energies close to the band gap in the whole FBZ and SBZ along the [001] direction are displayed in Fig.~1. The highest-energy states in the valence band are localized close to the $\Gamma$ point of the FBZ (the $\Gamma$ valley), whereas, the conduction-band states have their energy minimum at $k_z=K_0$, which is close to the $X$ point of the FBZ (the $\Delta$ valley). The parameters of the model were fitted such that they reproduce the dispersion relations in the full FBZ well. \cite{rideau2006} However, such a parametrization fails to produce the correct symmetry of the bands  with respect to the FBZ boundary as demonstrated by the solid lines in Fig.~1. For $k_z$ beyond the $X$ point the dispersion relations of all bands should be mirror symmetric to the dispersion relations left of the $X$ point, \cite{tadic2004,tadic2011} such as the ones shown by the dashed lines in Fig.~1. More specifically, a valley labeled by $\Delta^\prime$ in Fig.~1 should appear in the ground conduction band in the SBZ. However, this important detail is missing in the 30-band model. Rather, the energy of this band steeply increases with $k_z$ in the SBZ, and instead of the $\Delta^\prime$ valley, there exists  a valley of an upper conduction band, labeled by $\Theta$ in Fig.~1, which is just $80$ meV above the $\Delta$ valley. Also, it is located close to the FBZ boundary, therefore it can have an important contribution to low-energy conduction band states in the quantum well. Note that the symmetries of the two bands differ: The ground conduction band mainly has the $\Gamma_{15}$ zone-center symmetry, whereas the upper conduction band has the combined $\Gamma_{25u}+\Gamma_{2l}$ symmetry.

In addition to the lack of the symmetry of the conduction bands, the valence-band dispersion relations enter the band gap for large wave vectors, which is also shown in Fig.~1. Consequently, for a given energy $E<0$, there exists an additional wave vector outside the FBZ. It was demonstrated that, for quantum wells based on direct band-gap semiconductors and using the eight-band {\bf k}$\cdot${\bf p} Hamiltonian, these high-$k$ bulk states, which are degenerate with the low-$k$ bulk states produce spurious states in semiconductor quantum wells.\cite{yang2005} As we will see, the incorrect dispersions of the energy bands shown in Fig.~1 will have severe effects on the numerical calculations of the quantum-well states.

\section{The quantum well states}

For the hard-wall confinement potential, the quantum-well states are obtained by solving the equation,
\begin{equation}
    H_{30}\Xi=E\Xi.
\label{h30}
\end{equation}
Here, $H_{30}$ denotes the 30-band {\bf k}$\cdot${\bf p} Hamiltonian, which was introduced in Ref.~\onlinecite{rideau2006}, and $\Xi$ is the 30-band envelope-function spinor,
\begin{equation}
    \Xi=\begin{bmatrix}
            \chi_1,\chi_2,\ldots,\chi_{30}
         \end{bmatrix}.
\end{equation}
$\chi_j(z)$ denotes an envelope function of the zone-center periodic part of the Bloch function $u_j({\bf r})$.\cite{rideau2006} The full wave function of the electron in the quantum well reads
\begin{equation}
    \eta_{k_x,k_y}({\bf r})=\exp{[\imath(k_{x} x+k_{y} y)]}
    \suma_{j=1}^{30} \chi_{j}(z)u_j({\bf
    r}).
\end{equation}

In order to satisfy the Dirichlet boundary conditions, a basis of standing waves is chosen.\cite{richard2005,cukaric2012} Furthermore, for the conduction band the range of wave vectors of the basis states is conveniently centered at $k_z=K_0$ (see Fig.~1),\cite{richard2005}
\begin{equation}
    \chi_{j}(z)=\exp(\imath K_0 z)\sqrt{\frac{2}{W}}\suma_{m=1}^{N}c_m^{(j)}\sin(m\pi z/W).
\label{chibasis}
\end{equation}
Here, $N$ denotes the order of the basis and $W$ is the well (simulation box) width. We note that the conduction band of silicon has two minima along the [001] direction, which occur at $K_0$ and $-K_0$. This leads to a double degeneracy, which, along with spin, gives rise to four-fold degenerate states in the silicon quantum wells. A valley-splitting phenomenon breaks this degeneracy,\cite{ando1982} but this is a small effect due to both the inversion symmetry of the confining potential and the large separation between the equivalent $\Delta$ valleys at the $K_0$ and $-K_0$ points. Therefore, it is discarded in our calculations.

The dominant component of the envelope-function spinor $\chi_d=\chi_j$ is determined according to the criterion that it has the largest $C_j=\langle\chi_j\vert\chi_j\rangle$ out of 30 envelope functions which are the solutions of Eq.~(\ref{h30}).

\section{Results and discussion}

\subsection{The origin of spurious states}

\begin{figure}
        \begin{center}
       \includegraphics[width=8cm]{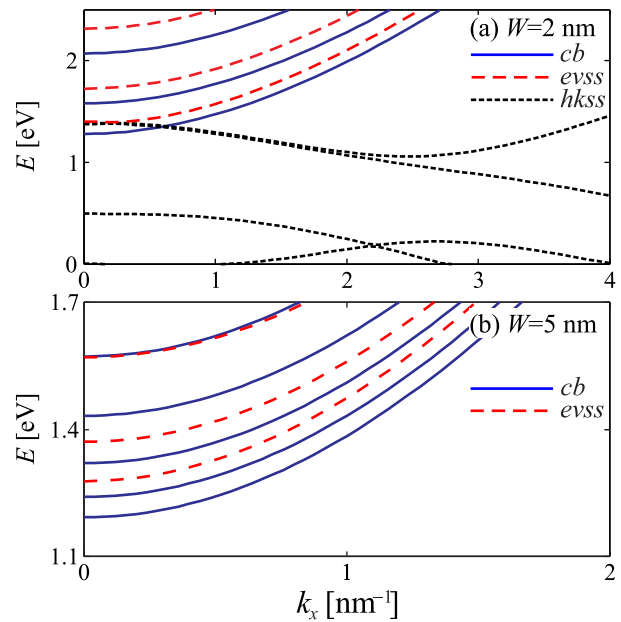}
       \caption{\label{fig2}(Color online) The dispersion relations of the subbands above the valence band top for the: (a) $W=2$ nm, (b) $W=5$ nm wide silicon quantum well. The solid lines denote the regular sub-bands, whereas, the dashed lines denote the spurious solutions. The basis size is $N=7$.}
    \end{center}
\end{figure}

The obtained spectra for quantum wells with thicknesses of $W=2$ nm and $W=5$ nm are shown in Figs.~\ref{fig2}(a) and 2(b), respectively. To obtain these results, the order of the basis $N$ in Eq.~(\ref{chibasis}) is chosen such that the ground-state energy is converged up to an accuracy of 1 meV. We found $N=7$ satisfies the convergence criteria. However, this leads to the presence of basis states outside the FBZ and inside the SBZ. For the conduction-band states (cbss) in the $W=2$ nm wide quantum well, out of the seven basis functions, just a single basis state belongs to the FBZ. It is a cumbersome detail related to the small separation of the $\Delta$ valley from the $X$ point. Furthermore, the $\Gamma_{15}$ states around the $\Delta$ point are expected to mostly contribute to the low-energy conduction-band states in the quantum well. However, the extra $\Theta$ valley is close in energy to the $\Delta$ valley, hence some quantum-well states will  be mainly $\Gamma_{25u}+\Gamma_{2l}$ like. Not all states shown in Figs.~\ref{fig2}(a) and 2(b) are physically relevant solutions, i.e., some spurious states are found in the energy spectrum. These states are denoted by dashed lines and are classified into two types as explained below.

Next we look at the localization of the electron in a few states of the $W=5$ nm wide quantum well for $k_x=k_y=0$ as shown in Fig.~3. In order to find both types of spurious solutions in the energy spectrum, we increased the basis size to $N = 15$. The probability densities of these states are displayed in the left panels [Figs.~3(a)-(c)], whereas the right panels [Figs.~3(d)-3(f)] show  $\tilde{\chi}_d(z)=\chi_d(z)/\exp(\imath K_0 z)$. The probability density of the cb ground state and the dominant envelope function shown in Figs.~3(a) and 3(d) resemble those of the ground state of the infinite rectangular quantum well according to the single-band model. It implies that the dominant contribution to the cb ground state arises from the bulk states whose wave vectors are around the $K_0$ point.

On the other hand, the spurious solution with an energy of 453 meV has both a highly oscillatory probability density and the dominant envelope function as depicted in Figs.~3(b) and 3(e), respectively. The dominant envelope function is almost regularly periodic with a period of 0.7 nm, therefore, it is mainly composed of the bulk state with the wave vector $k_z=2\pi/(0.7\text{\ nm})+K_0\approx 18$ ${\rm nm}^{-1}$. Such a high-$k$ value is outside the FBZ where the 30-band model previously was demonstrated to fail. Therefore, such states are named {\it high-$k$ spurious solutions}, and are abbreviated by hksss (denoted by the short dashed lines in Fig.~2). They are found in both the conduction and the valence bands.

\begin{figure}
        \begin{center}
       \includegraphics[width=8cm]{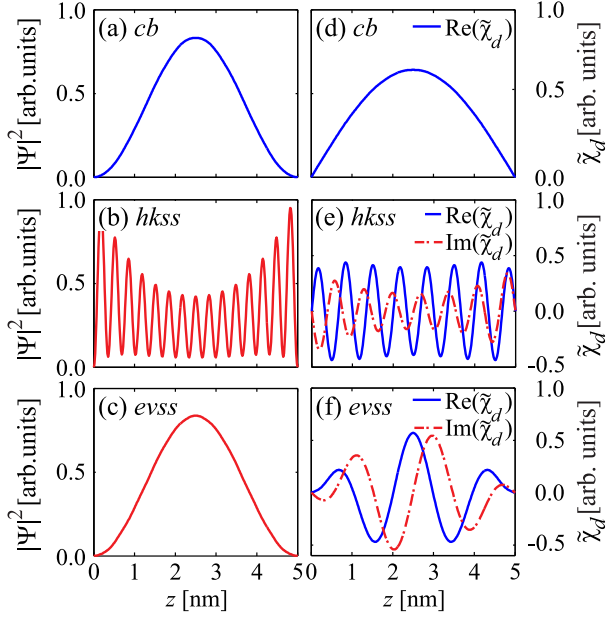}
       \caption{\label{fig3}(Color online) (Left panel) The probability density as function of $z$ for a few states in the silicon quantum well of width $W=5$ nm: (a) the cb ground state, with energy $E=1193$ meV; (b) the high-$k$ spurious solution (hkss) with energy $E=453$ meV, (c) the extra valley spurious solution (evss), whose energy equals $E=1275$ meV. (Right panel) The dominant components of the envelope function spinors of the states shown in the left panel are divided by $\exp(\imath K_0 z)$: (d) cb, (e) hkss,  and (f) evss. The imaginary part in the cb state is 2 orders of magnitude smaller than the real part. The basis size is $N=15$.}
    \end{center}
\end{figure}

Figures 3(c) and 3(f) display the state, whose energy is 1275 meV, which sets in between the cb and hkss states shown in Fig.~3. As a matter of fact, its probability density shown in Fig.~3(c) resembles the cb ground state shown in Figs.~3(a) and 3(d) and, therefore, could solely indicate that the state is a regular one. However, the dominant component of the envelope function spinor, shown in Fig.~3(f), is more oscillatory than $\tilde{\chi}_d$ displayed in Fig.~3(d). Yet, these oscillations are less regular and of the larger period than for the hkss state [compare Figs.~3(e) and 3(f)]. Nevertheless, they are composed of the wave vectors outside the FBZ [$k_z=2\pi/(1.7\text{\ nm})+K_0\approx 13$ ${\rm nm}^{-1}$], and are mainly contributed by the bulk states of the $\Theta$ valley. Therefore, such states are spurious but of another type, which are named evss. They are denoted by the long dashed lines in Fig.~2 and are found only in the conduction band.

In order to further illustrate the origin of the states displayed in Fig.~3, in Fig.~4, we show the corresponding distributions of the probability over the different components of the envelope-function spinor $\langle\chi_j\vert\chi_j\rangle$.  Figure 4(a) shows $\langle\chi_j\vert\chi_j\rangle$'s for the electron ground state and demonstrates that this state mainly is composed of the $\Gamma_{15}$ zone-center states. However, it has a large contribution from the $\Gamma_{1u}$ band.\cite{rideau2006} This result is consistent with the approximate two-band model proposed in Ref.~\onlinecite{michelini2011}. On the other hand, the main contribution to the hkss of Fig.~3(b) comes from the $\Gamma_{25l}$ states as shown in Fig.~4(b), and the largest $\langle\chi_j\vert\chi_j\rangle$ in the evss displayed in Fig.~3(c) belongs to the $\Gamma_{25u}$ and $\Gamma_{2l}$ bands, as Fig.~4(c) shows. The latter two bands mainly form the $\Theta$ valley just outside the FBZ (see Fig.~1), which is an artifact in the SBZ, thus such states are classified as spurious.

\begin{figure}
        \begin{center}
       \includegraphics[width=8cm]{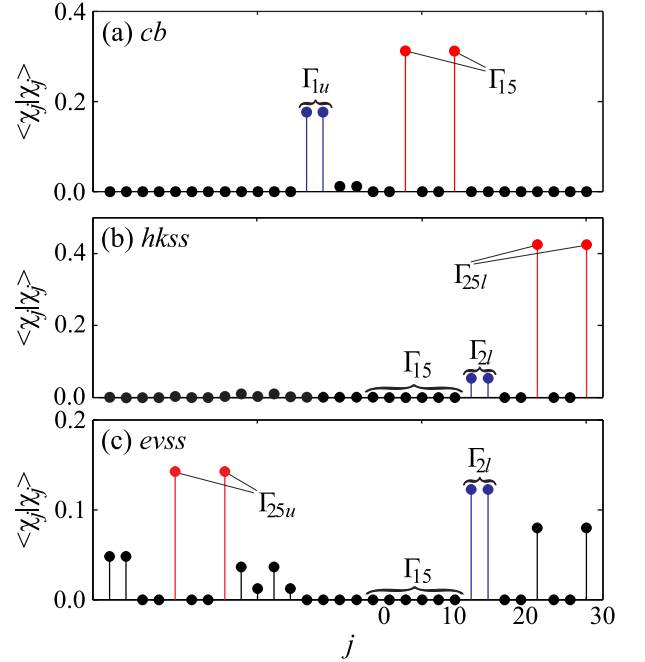}
       \caption{\label{fig4}(Color online) The envelope function spinor dominant component of the states shown in Fig.~3: (a) cb, (b) hkss,  and (c) evss.}
    \end{center}
\end{figure}

\subsection{The spurious solutions removal}

We developed a scheme to automatically remove both types of spurious solutions. It is based on the following observations. In addition to contributions of different zone-center states to quantum-well states, which were illustrated in Fig.~4, the absolute value of the expansion coefficients $\vert c_m^{(j)}\vert$ is found to be an important figure of merit for classifying  quantum-well states as regular and spurious ones. We checked the distributions of $\langle\chi_j\vert\chi_j\rangle$ over $j$ and $\vert c_m^{(j)}\vert$ over both $j$ and $m$ and were able to formulate the set of empirical rules for extracting a few (three to five) low-energy spurious solutions from the {\it conduction-band } spectrum of the quantum well. The regular states in the conduction band are found to mainly originate from the $\Gamma_{15}$ band. We label the regular conduction-band states by the counter $n$. Furthermore, $\chi_{\Gamma_{15}}$ envelope functions were found to mostly be composed of the low-$m$ basis states. For example, the electron ground state for the range from $W=2$ nm to $W=20$ nm is found to mainly be composed of the $m=1$ basis function. Furthermore, the $m$ values of the expansion coefficients with the largest magnitudes in the conduction-band states $n$ and $n+1$ are found to differ by not more than unity. Also, the quantum-well states, whose dominant envelope function $\chi_d$ is due to bulk states different from $\Gamma_{15}$, are found to be composed dominantly of the basis functions with wave vectors outside the FBZ. Therefore, they are spurious in origin, and may be of the hkss or evss type.

The proposed {\it modus operandi} is as follows. The calculation starts by choosing the value of the order of the computational basis $N$ to achieve a reasonable energy accuracy, as previously explained. The Hamiltonian is then diagonalized, and the envelope functions with the largest $\langle\chi_j\vert\chi_j\rangle$'s are selected for all the computed states. The index of the dominant envelope function is labeled by $j_{max}$. Furthermore, for the determined $j_{max}$, the largest expansion coefficient $\vert c_m^{(j_{max})}\vert$ is found and is labeled by $m=m_{max}$. The $m_{max}$ value will be compared with $\tilde{m}$, which is the reference value of $m_{max}$, and is set to unity when the procedure starts. The procedure for eliminating the spurious solutions from the spectrum of the conduction-band states reads:
\begin{enumerate}
\item{Set the reference value of the maximal index of the dominant basis function to $\tilde{m}=1$; set the number of the regular states to $n=0$.}

\item{The composition of all the states from zero energy onward is determined; a state with the largest contribution of the $\Gamma_{15}$ band is selected for further consideration;}

\item{For the selected state, $c_{m_{max}}^{(j_{max})}$ is determined.}

\item{If $m_{max}>\tilde{m}$ the state is classified as spurious.}

\item{If $m_{max}=\tilde{m}$, increase $\tilde{m}$ by 1, i.e., $\tilde{m}=\tilde{m}+1$; such a state is classified as regular, thus $n=n+1$.}

\item{If $m_{max}<\tilde{m}$, the state is classified as a regular state, and therefore, $n=n+1$.}

\item{Go back to step 2 to proceed with checking the other states.}
\end{enumerate}

Note that no regular state is misclassified by this procedure. In other words, we found that the states which do not have the dominant $\Gamma_{15}$ component are dominated by standing waves with wave vectors outside the FBZ.  However, the proposed procedure may be applied to only remove a few lowest-energy spurious states, which is three to five for $W$ ranging from 2 to 5 nm. Mixing between the $\Gamma_{15}$ and  the $\Gamma_{25u}$+$\Gamma_{2l}$ zone-center states, which form the $\Delta$ and $\Theta$ valleys, respectively, becomes larger when the electron energy increases, and the explained algorithm cannot be adopted. Furthermore, the proposed algorithm cannot be applied to thin quantum wells, and $W=2$ nm was found to be a practical lower limit. For quantum wells thinner than approximately 2 nm, convergence of the electron energy to within 1 meV is not reachable.

\begin{figure}
        \begin{center}
       \includegraphics[width=8cm]{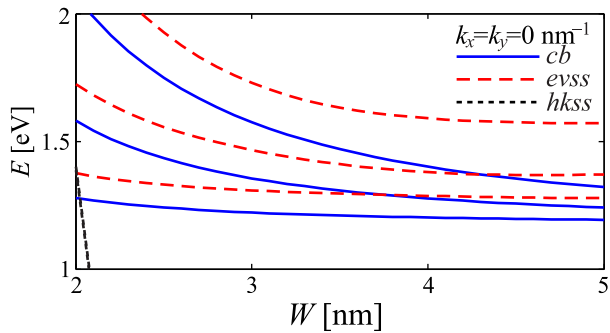}
       \caption{\label{fig5}(Color online) The dependence of the regular energy levels (solid lines), evsss (long dashed lines), and hksss (short dashed lines) for $k_x=k_y=0$ on the quantum-well width for the $N=7$ basis size.}
    \end{center}
\end{figure}

Both the hksss and evsss are found to exist in the range of $W$ from 2 to 5 nm for the chosen basis size. When $W$ increases, the evss energies cross the energies of the regular states, as shown in Fig.~5. Similar to Fig. 3, the $(k_x,k_y)=(0,0)$ states are shown in this figure. The number of the regular states, whose energies are lower than the lowest-energy evss increases with $W$. Therefore, when $W$ tends to infinity, which is the bulk silicon case, all the regular states will be below all evsss. In the energy range displayed in Fig.~5, only two hksss are above the ground conduction-band states for $W=2$ nm, and their energies sharply decrease with $W$, such that already for $W=2.1$ nm these hksss enter the band gap where they can easily be recognized and removed from the energy spectrum.

As discussed, Fig.~2(a) shows the dispersion relations of the sub-bands which have energies close to the conduction-band bottom in the 2-nm-wide well. Because of band folding the minimum of the conduction band is at $k_x=k_y=0$. Some spurious solutions evidently are found in the band gap, and all of them are of the high-$k$ type and are, therefore, easily removed. On the other hand, a few evsss are found in the conduction band, whose dispersion relations appear to be similar to the dispersion relations of the regular states. It is because $evss's$ are formed out of the states of the $\Theta$ valley, which is similar to the real conduction band states which mainly arise from the states of the $\Delta$ valley. In other words, the bulk states of both the real states and the evsss do not exhibit appreciable band mixing.

\begin{figure}
        \begin{center}
       \includegraphics[width=8cm]{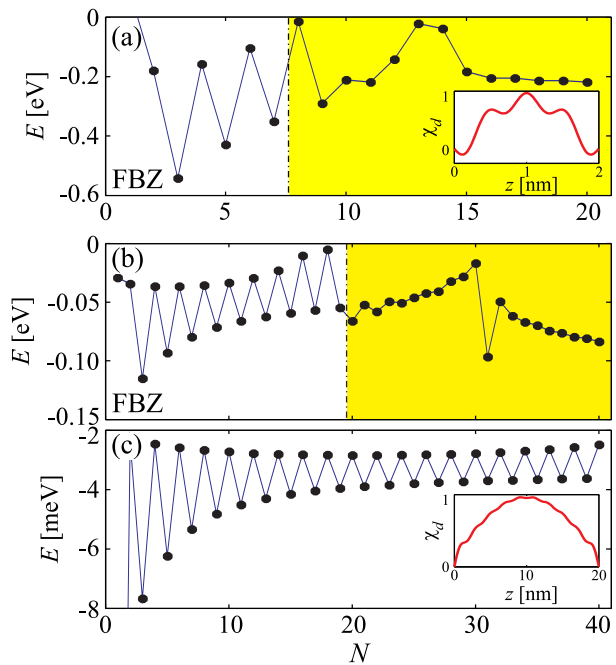}
       \caption{\label{fig6}(Color online) The hole ground state energy as function of the basis size $N$ for (a) $W=2$ nm, (b) $W=5$ nm, (c) and $W=20$ nm. The yellow color displays the area outside the FBZ. The insets show the dominant component of the hole ground-state envelope-function spinor.}
    \end{center}
\end{figure}

\subsection{The hole states}

Let us now consider the hole states. The presented procedure can also be adopted to remove the spurious solutions in the energy range of the valence band, except that the real valence-band states are found to mainly be composed of the $\Gamma_{25l}$ zone-center states. But, in addition to the spurious solutions, the hole states in the silicon quantum well suffer from an instability in the calculation with respect to the basis order as Fig.~6 demonstrates. Notice that the hole ground-state energy level oscillates with the size of the basis. The amplitude of the oscillations can be as large as 100 meV, and its value decreases when the well width increases, as Figs.~6(a)-6(c) show for $W=2$, 5, and 20 nm, respectively.

These zigzag-shaped convergence can be explained as follows. First, note that the diagonal elements of the 30-band Hamiltonian are equal to the kinetic-energy term for a free electron. Therefore, without band mixing, the dispersion relations of all bands in silicon are concave. The curvature of the valence band alters sign through the band mixing. The analyzed silicon quantum well is symmetric, and for $k_x=k_y=0$, the envelope functions are classified strictly with respect to inversion of the $z$ coordinate as even or odd. The dominant component of the hole ground state is even, therefore, it is composed of the $m=1,3,5,\ldots$ basis states. These basis states are dominantly coupled with the odd ($m=2,4,6,\ldots$) basis functions by the off-diagonal terms which are proportional to $k_z$. It is obvious from the form of the 30-band Hamiltonian given in Ref.~\onlinecite{rideau2006} that the finite overlap between the even and the odd envelope-function spinor components leads to a change in the sign of curvature of the sub-band dispersion relation.

To further illustrate the zigzag variation in the hole eigenstates observed in Fig.~6, we focus on a result obtained with a basis of size $N$ and one with size $N+1$, where $N$ is an odd number. The extra basis function in the $N+1$ basis is an odd function. Because the slope of this $(N+1)$th basis function is largest close to the boundary where the oscillatory $N$th basis function reaches its maximum, the value of the matrix element between the two states can be large and, therefore, can substantially modify the eigenenergy value. For odd $N$, the basis is, in fact, not effective in establishing the appropriate curvature of the quantum-well sub-bands. Also, the dominant envelope function in the spinor of the ground hole state has extra zeros close to the well boundary as the inset in Fig.~6(a) shows for $N=7$ and the $W=2$ nm quantum well. However, if $N$ is an even number, the dominant envelope function of the ground hole state becomes less oscillatory, and the extra zeros of the envelope function do not exist as the inset in Fig.~6(c) demonstrates for $N=20$ and the $W=20$ nm wide quantum well.

The demonstrated instability of the valence-band solutions with the size of the basis essentially is a consequence of the inappropriate curvature of the hole states as modeled by the diagonal terms of the 30-band model. Such problems do not exist in the six-band model where the sign of curvature of the valence-band dispersion relation is appropriate even if modeled by only the diagonal terms. The problem cannot be solved by increasing the size of the basis, i.e., by taking states outside the FBZ as shown in Figs.~6(a) and (b) into account. In fact, it arises from the need of the envelope function to drop exactly to zero at the boundary.

\subsection{The case of the finite band offset}

In order to explore how the assumption of the infinite barrier affects the stability of the hole states calculations, we extend our analysis to the case of a finite-depth ${\rm Si}/{\rm SiO}_2$ quantum well. The valence-band offset in ${\rm Si}/{\rm SiO}_2$ systems has been found to amount to 4.5 eV.\cite{bersch2008,alay1997} However, no values for the parameters of the 30-band model have been extracted, therefore, they are assumed to be equal to the parameters of silicon, except for the value of the band gap at the $\Gamma$ point, which equals 8.9 eV.\cite{bersch2008,tan2005} In these calculations, we assume that the Si well of width $D$ is centrally positioned with respect to the simulation box, whose width is denoted by $W$. As an example, we assume that the Si well is $D=5$ nm wide and choose $k_x=0$ nm$^{-1}$ and $k_y=0$ nm$^{-1}$.

\begin{figure}
        \begin{center}
       \includegraphics[width=8cm]{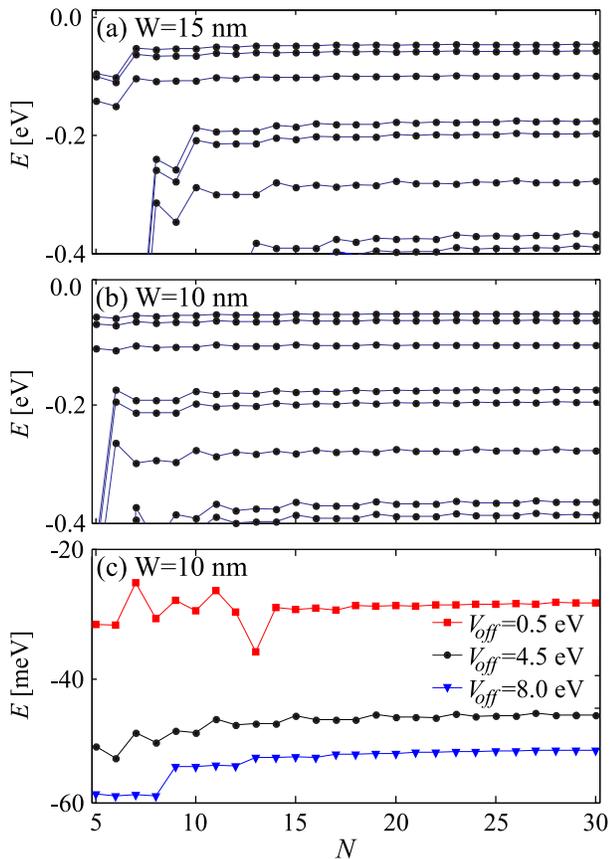}
       \caption{\label{fig7}(Color online) A few hole states in the $D=5$ nm wide quantum well of finite depth as a function of the basis size $N$ for (a) $W=15$ nm, and (b) $W=10$ nm. (c) Convergence of the hole energy levels with $N$ for three values of the valence-band offset.}
    \end{center}
\end{figure}

The obtained eight highest hole energy levels for $W=15$ nm and $W=10$ nm as functions of the basis order $N$ are shown in Figs.~7(a) and 7(b), respectively. Quite interestingly, the oscillations previously found for the case of the infinite quantum well do not take place when the valence-band offset is finite. For both values of $W$, the results improve by increasing $N$, and as expected, the smallest basis is needed to compute the ground state. Furthermore, no big change is observed when the size of the simulation box decreases from $W=15$ nm to $W=10$ nm, except that a slightly larger basis is needed when the simulation box is wider. Therefore, allowing the envelope functions to exponentially decay stabilizes the energy-level dependence on $N$.  It confirms our previous claim that the steep descents of the envelope functions near the quantum-well boundaries cause numerical instabilities with the hard-wall potential shown in Fig.~6.  Furthermore, as Fig.~7(c) demonstrates, when $N$ is sufficiently large ($N\geq 14$), we found that quite reliable results are produced irrespective of the value of the valence-band offset. In this figure, the ground-state energies in two unrealistic cases, $V_{off}=0.5$ eV  and $V_{off}=8$ eV, are shown along with the ground state for $V_{off}=4.5$ eV, which was previously shown in  Fig.~7(b). This figure demonstrates that, if the envelope function is allowed to exponentially decay to zero inside the barrier, computation of quantum-well states becomes quite stable with respect to the number of basis functions. Even for a valence-band offset as large as 8 eV, the convergence of the hole ground-state energy level towards the numerically exact value is found to be quite steady, and only $N=13$ basis states are needed to produce the energy value with a negligible error.

The value of the valence-band offset $V_{off}=4.5$ eV is large such that the envelope functions decay fast in the barrier. Hence, the energy of the hole ground state is almost constant for $W>6$ nm as Fig.~8(a) shows for $N=20$. The energies of the other states depend similarly on $W$. However, some of them clearly exhibit oscillations. The reason is as follows. Since lower states of the valence band are more oscillatory, we need a broader $k$-interval than for the ground state to accurately describe them. With increasing box size (and fixed basis size $N$), we narrow the covered $k$ space (since $k\sim1/W$), so these low states are not described accurately. Nevertheless, we need a wider box for these states than for the ground state. When the difference between $W$ and $D$ is not large, the confining potential is like in the infinite quantum well. Consequently, the results for the highest-energy states become quite unstable when $N$ varies as Fig.~6 previously showed.

\begin{figure}
        \begin{center}
       \includegraphics[width=8cm]{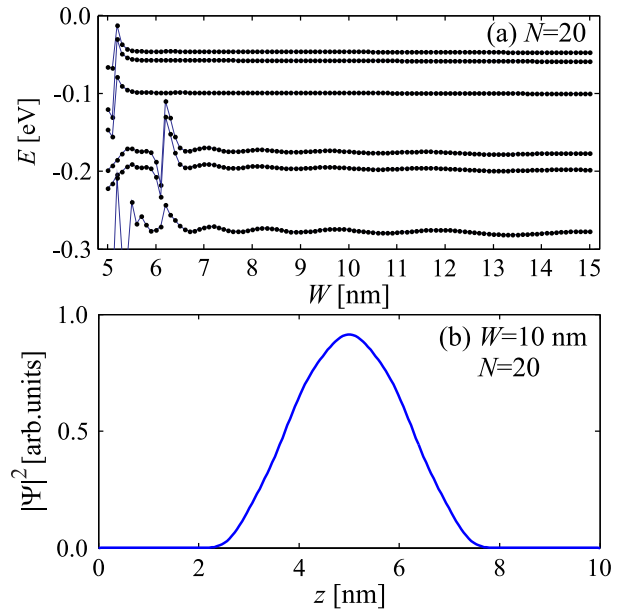}
       \caption{\label{fig8}(Color online) (a) The hole ground state energy in the $D=5$ nm wide quantum well of finite depth as a function of the box size for $N=20$. (b) The probability density of the hole ground state for $W=10$ nm and $N=20$.}
    \end{center}
\end{figure}

Figure~8(a) indicates that, if one is interested in computing only three highest energy states, even the choice $W=6$ nm produces a good result. Varying width and depth of the quantum well might modify this finding, and could depend on the values of the material parameters. But Fig.~8(b) shows that the probability density of the hole ground state is quite confined inside the well (which ranges between $2.5\text{\ nm}\leq z\leq 7.5\text{\ nm}$). It accounts for why the energy of the ground state in Fig.~8(a) does not vary much for $W>6$ nm. Moreover, we found that, even for a small barrier width, the highest-energy states can be computed quite accurately, and the accuracy of the calculation of the lower energy states can be improved by increasing the basis order. Hence, thin silicon layers embedded between thick barriers can be modeled accurately by the employed 30-band theory, providing the width of the simulation box and the basis size is large enough. We note that, for finite band offsets, Richard {\it et al.} previously employed 40 basis functions in the 30-nm-wide box to compute the hole states in the ${\rm Ge}/{\rm SiGe}$ quantum well with 1-meV accuracy.\cite{richard2005}

\section{Conclusion}

We used a basis consisting of standing waves within the 30-band {\bf k}$\cdot${\bf p} model to solve the electronic structure of a ${\rm Si}/{\rm SiO}_2$ quantum well grown in the [001] direction. For the assumed infinite potential steps at the well boundaries, we found that numerous spurious solutions are present in the computed electron and hole spectra. These spurious states are classified into two categories: the high-$k$ states which arise from the contribution of the states outside the first Brillouin zone and the extra-valley spurious states which arise from the spurious valley outside the first Brillouin zone. The missing symmetry of the conduction band in bulk silicon as modeled by the 30-band {\bf k}$\cdot${\bf p} Hamiltonian is found to be the cause of the extra-valley spurious states in the conduction band. Furthermore, we devised a procedure which is able to remove the low-energy spurious states from both the conduction-band and the valence-band energy spectra. The latter is found to exhibit instabilities due to a peculiar band mixing and the specific boundary conditions, when the order of the employed basis varies. This failure of the 30-band {\bf k}$\cdot${\bf p} model might heuristically be accounted for by a large difference in the electron confinement in the hard-wall silicon quantum well and the silicon bulk. However, if the hard-wall confinement is made softer the deficiencies in the 30-band {\bf k}$\cdot${\bf p} approach are found to disappear for the adequately chosen size of the simulation box and the basis order. Furthermore, the choice of the numerical method is not relevant for the demonstrated instability of the hole states, i.e., we found that it also exists if the finite-difference or finite-element methods are adopted to solve the 30-band eigenvalue problem.

\section*{ACKNOWLEDGMENTS}

This work was supported by the Ministry of
Education, Science, and Technological Development of Serbia, the Belgian Science Policy (IAP), the Flemish fund for Scientific Research (FWO-Vl), and the Methusalem programme of the Flemish government.

\end{document}